\documentstyle[multicol,aps,psfig]{revtex}
\begin{document}
\title{{\small Phys. Rev. E., Vol. 50, 2654--2659 (1994).}\\[0.5cm]
Synchronization effects in the dynamical behavior of elevators}
\author{Thorsten P\"oschel\cite{bylinetp} and 
               Jason A.~C.~Gallas\cite{bylinejg}}
\address{H\"ochstsleistungsrechenzentrum, 
    Forschungszentrum J\"ulich, D-52425 J\"ulich, Germany}
\date{\today}
\maketitle
\begin{abstract}
We simulate the dynamical behavior of $M$ elevators
serving  $N$ floors of a building in which a Poisson distribution of
persons call elevators. Our simulation reproduces the
jamming effect typically seen  in large buildings  when
a large number of persons decide to leave the building
simultaneously.
The collective behavior of the elevators
involves  characteristics similar to those observed 
in systems of coupled oscillators. In addition,
there is  an apparently  rule-free critical population density 
above which elevators start to arrive
{\it synchronously\/} at the ground floor.
\end{abstract}
\setcounter{topnumber}{3}
\renewcommand{\topfraction}{1.0}
\setcounter{topnumber}{5}
\setcounter{bottomnumber}{5}
\setcounter{totalnumber}{5}
\renewcommand{\topfraction}{1}
\renewcommand{\bottomfraction}{1}
\renewcommand{\textfraction}{0}
\begin{multicols}{2} 
The macroscopic effect of collective synchronization is a remarkable
phenomenon occurring in several fields and is presently
a topic of very active research. Effects of collective synchronization
can be observed in situations as diverse as, for example,  the
dynamics of sliding charge-density waves\cite{1.},  the phase-locking
of relativistic magnetrons\cite{2.}, and many other situations of
interest in physics and engineering\cite{3.}. In addition,
as early observed by Winfree\cite{4.}, synchronization is  a
phenomenon that also occurs in virtually all levels of biological
organization. Interesting examples of mutual synchronization
involving biological phenomena are epileptic seizures in the brain,
electric synchrony among cardiac pacemaker cells, flashing synchrony
among swarms of fireflies, crickets chirping in unison, synchronization
of menstrual cycles in groups of women, and several others\cite{5.}.
A common way of investigating the dynamics of
such collections of oscillators is by assuming a relatively weak
coupling $K$ among them together with a random distribution of 
eigenfrequencies. Increasing coupling from zero
one finds  incoherent collective
behavior up to a  critical threshold after which the system
starts to behave in a state displaying relatively high degrees of
synchronization among the individual oscillators composing the system.
A convenient  way of dealing analytically with
several coupled oscillators is via a model proposed  by
Kuramoto\cite{6.}, who assumed the dynamics to be ruled by the
set of equations
$$\dot\theta_i = \omega_i +{K\over N} \sum_{j=1}^N 
           \sin(\theta_j-\theta_i),  \qquad i=1, 2,\dots,N. \eqno(1)$$
In this equation N is the number of coupled oscillators while
$\theta_i$ and  $\omega_i$ are their phase and
natural frequency, respectively.
Recent progress in the understanding of the dynamical behavior of
systems containing relatively few degrees of freedom along with
a strong motivation to understand nontrivial temporal and spatial
structures has led to much interest in models for spatially
extended systems\cite{7.}. In several of these investigations, not
only the somewhat more traditional mathematical approach of using
sets of either ordinary or partial differential equations has been
used, but new aspects have been uncovered by also using 
coupled map lattices and cellular automata\cite{7.}\cite{8.}\cite{9.}.

The purpose of this paper is to report synchronization effects
observed in a computer simulation of the dynamical behavior of $M$ elevators
serving $N$ floors of a building. 
Specifically, we report here the dynamics of the {\it Feierabend effect},
that is, the jamming effect occurring at a certain time when a large number
of occupants of a building decide to leave it nearly simultaneously.
Macroscopic synchronization effects between elevators were observed 
under a number of different conditions, all of them
variations around a simple
set of ``working rules'' assumed to define the dynamical behavior
of the elevators. 
These rules were chosen in such a way  to provide a reasonably fair 
representation of what one usually gets 
from not very sophisticated, not computerized elevators.
The  definition a set of rules controlling the behavior of elevators 
is not a completely trivial task
since the movement of elevators depends on a quite large number of factors.
To start, while one might easily agree on a working definition for
{\it small\/} and {\it large\/} buildings, it is  hard to
differentiate unambiguously the ``transition'' from
small to large buildings. This question is relevant because
elevator rules good  for small buildings 
are not practical for higher buildings and vice-versa. 
In higher buildings, it is common practice to 
divide the total number of elevators  into smaller
groups in order to ensure good service for all floors. This division
is not unique: one may use some elevators to serve only  odd-numbered
floors while the others serve even-numbered floors; or one may decide that
a number of elevators will serve only the first, say, 10 floors,
the next will serve floors 10 to 20, and so on. It is not difficult to
imagine a number of other possible combinations.
In addition to all this, there are a number of specific details
that need to be defined in order to optimize the process,
i.e.~to minimize the time needed for a person to move between floors
and to minimize energy consumption.

In what follows we will assume the
ground floor to be the only possible way to exit the building and that
all calls for elevators are from passengers wanting to exit by going
to this floor. This choice reflects closely the phenomena that we want
to consider. 
Each elevator can carry no more than 20 passengers simultaneously.
The specific working rules
of the  elevators are  assumed to be as follows.
Whenever  free elevators are available, every call originating from a
floor from which there were no previous unresolved calls  will set 
the closest free
elevator in motion. This will occur regardless of the relative
position of the floor originating the call and of  any  pre-existing
movement of elevators. 
After fulfilling the
original call that set it moving, every elevator will always try to
attend all other calls appearing in the floors lying below it, namely,
it will stop at all levels lying below.
While in practice one knows that an elevator already full cannot accept
new passengers, {\it real\/} elevators are frequently  not able to 
recognize and use the information that they are full to 
avoid stopping unnecessarily at lower floors. 
We simulate both situations: ``naive'' elevators, elevators which
even after being full will continue to stop for every call from below,
and ``smart'' elevators, elevators that once full will ignore 
all subsequent calls, moving straight to the lowest floor.
Empty elevators arriving at empty floors, i.e.~floors that have 
already been served, or on the ground floor
will stop and either wait for a new call or move to floors
where there are passengers calling,  which could not be served
earlier.

The updating of the dynamics is done at regularly spaced  time intervals,
controlled by the discrete ticking of a clock. 
At every clock-step we assume  $p_\mu(n_k)$ new  passengers 
to call elevators 
from the $k-$th floor, distributed according to a Poisson law
$$ p_\mu(n_k) = \mu^{n_k}{e^{-\mu}\over n_k!}, \eqno(2)$$
where $\mu$ is the Poisson parameter. At every clock-step the
location and movement of every elevator will also be updated. 
The speed of the elevators is one floor per time-step, either up or
down. Whenever an elevator stops in some floor, it remains in the
floor during a few clock units. This time is intended to represent an
average time consumed in opening the door, moving passengers and
closing door. In the present simulations we assumed this time to be
5 clock units.
 In absence of any new call,
elevators will preserve their previous state of movement.
Some elevators may stop, if they arrive on the first floor. 
Otherwise, elevators will react to new calls as described above.
To investigate the dynamical behavior as a function of $\mu$
we perform simulations over relatively long time intervals (typically
$t=5000$ clock steps) accumulating several quantities at every
clock-step, 
 {\it e.g.}~the total number of elevators that arrive at the ground floor,
 the number of times  that $L$ elevators arrive
 {\it simultaneously\/} at the ground level,
 the cumulative number of passengers transported,
 the number of passengers waiting in each queue, etc.

\begin{figure}[htbp]
\centerline{\psfig{figure=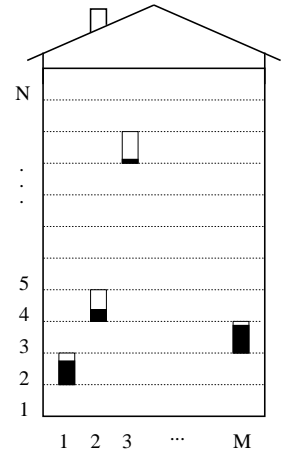,width=4cm}}
\caption{ Schematic representation of a typical building. The black
 shading
  of each elevator is proportional to the number of passengers that it
carries. 
}
\label{house}
\end{figure}

Based on these  assumptions we wrote a FORTRAN program to simulate the
dynamics of a few different buildings with several possible number of
elevators. Simulations were done over long time intervals and for
different values of the Poisson parameter $\mu$.
Apart from storing the aforementioned  relevant quantities describing 
the process, the program could also be run interactively, 
generating a  PostScript on-screen animation of the dynamics,
which could be easily video-recorded. 
An schematic snapshot of a video-frame is given in
Fig.~\ref{house}. 
The shading of each individual elevator is
proportional to the number of passengers that it carries.
Further (not shown in the schematic figure), the actual number of
passengers waiting on every floor was displayed  on the left of the
rectangle representing the building
by a line segment with length proportional to the population
density. Line segments on the right side indicated the 
{\it cumulative\/} number of passengers transported since the beginning
of every run.
The majority of runs was done for $M=10$ and $N=50$. For numerical
values of this order, 
the position, the population density inside every elevator (their
shading) and the population density outside every floor (line
segments) may be comfortably  updated on screen at every time step,
producing an actual movie of the dynamics.
In this way we are always able to visualize simultaneously the up-and-down
movement of all elevators and to monitor  the behavior of the
population  inside every elevator and on the several waiting queues. 
By monitoring the number of passengers in the waiting queues
and inside elevators, it was possible to recognize many
interesting effects such as, for example, when the number of elevators
was well adapted to the size of a building, to the population density
living in the building, etc. Simulation of a particular building
consisted essentially of ``playing the elevator game'' for different
values of $\mu$ and observing the dynamics on screen while cumulating
important numbers for later analysis.

\begin{figure}[htbp]
\centerline{~~~~~~~~~~~\psfig{figure=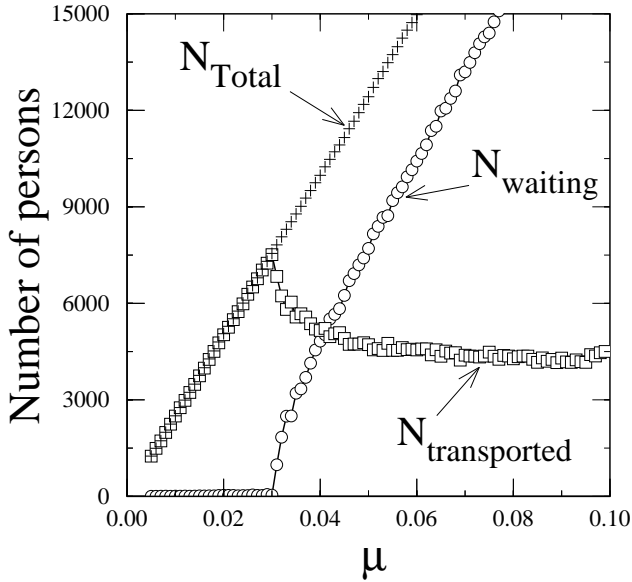,width=8cm}}
\vskip 1cm
\caption{ Accumulated population  density after 5000 time
steps. Above $\mu\simeq0.03$ the transport is not efficient anymore.}
\label{fig1}
\end{figure}

\begin{figure}[htbp]
\centerline{\psfig{figure=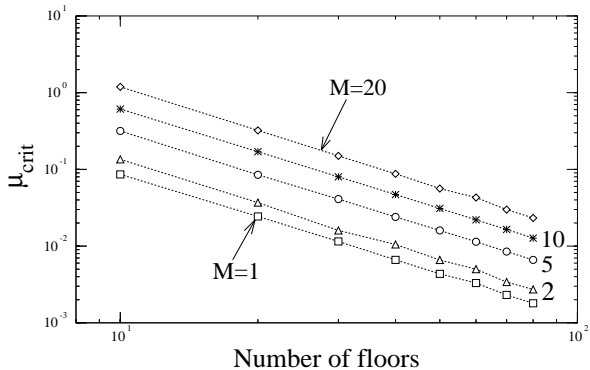,width=8cm}}
\caption{ Critical values of $\mu$ as functions of the number of
  floors and number of elevators. As seen in Fig.~2, beyond
  $\mu_{crit}$ the transport is not efficient anymore.
  The characteristic exponent underlying all curves is
   $\mu\simeq N^{-1.88}$.  }
\label{mucrit}
\end{figure}

Figure \ref{fig1} shows for $\mu$ between $0$ and $0.1$
the number of passengers  transported and waiting elevators after
5000 clock steps. One clearly sees the existence of an operational
``jamming threshold'': for $\mu$ slightly above the {\it critical\/} value
$\mu_c\simeq 0.03$ the 10 elevators are not able to cope with the
traffic anymore: the number of waiting passengers increases rapidly while
the number of passengers transported tends rapidly to the 
maximum limit afforded by the capacity of the total number of elevators.
Figure \ref{mucrit} shows  a plot the critical values of
$\mu$ obtained by simulation the dynamics of buildings containing
$10\leq N \leq80$ floors and served by either 
$M=1, 2, 5, 10$ or  $20$ elevators.

\begin{figure}[htbp]
\centerline{\psfig{file=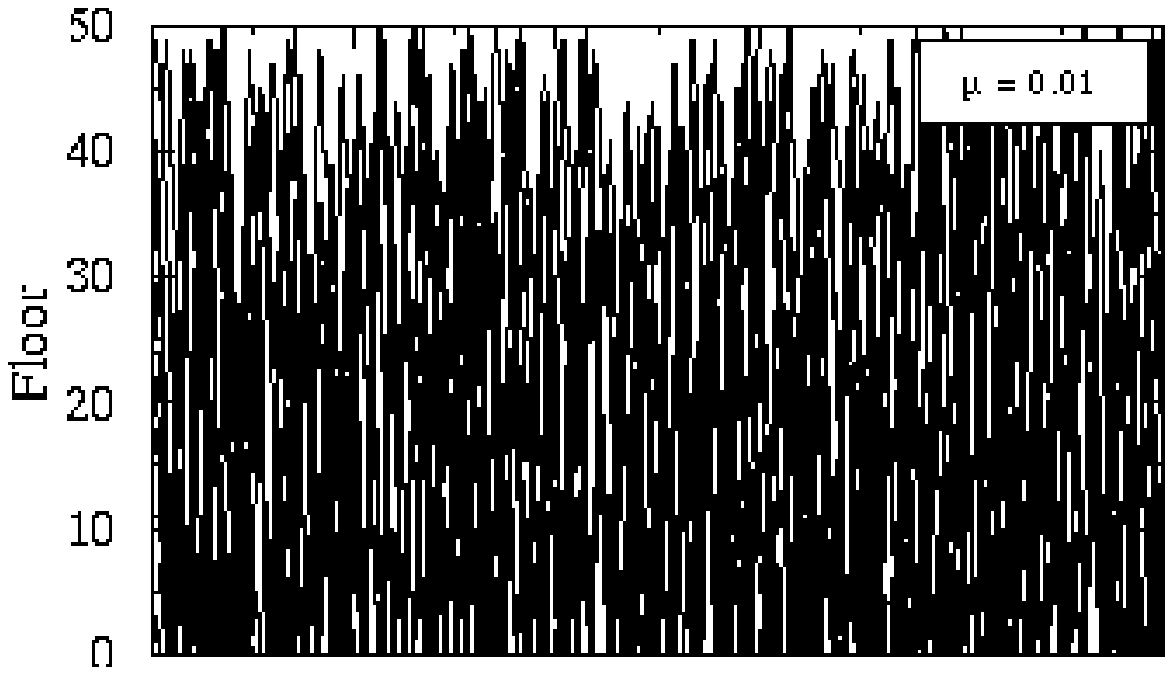,width=8cm}}
\centerline{\psfig{file=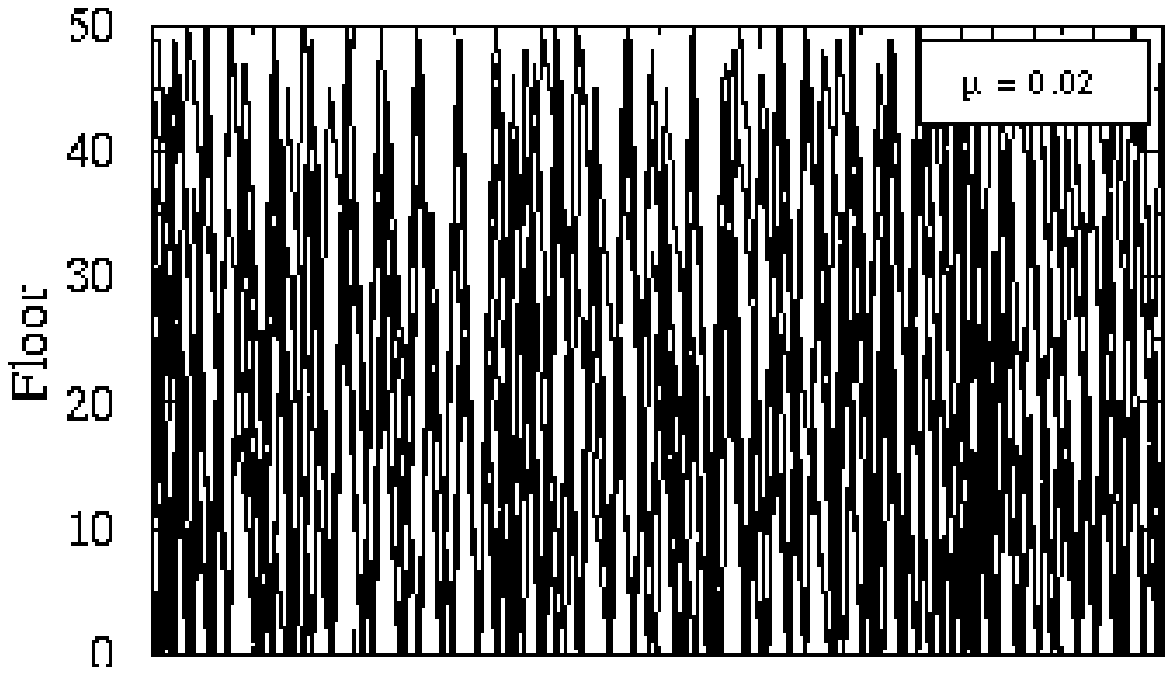,width=8cm}}
\centerline{\psfig{file=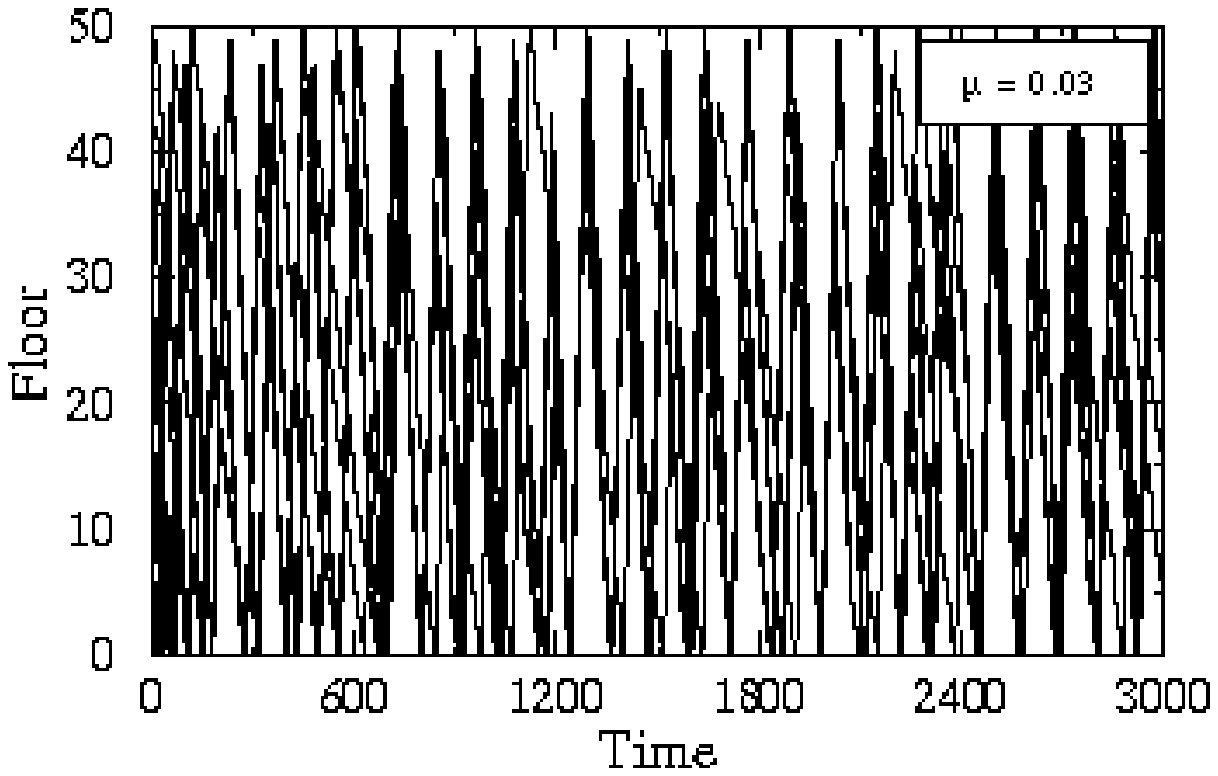,width=8cm}}
\caption{Up-and-down movement of 10 elevators  as a function
of time for three different Poisson distributions of calls.}
\label{fig2}
\end{figure}

As one sees from the graph, the behavior on a log-log plot may be very
well fitted by straight-lines.
By changing
the size of the building, the capacity of the elevators and/or the number
of elevators serving the building one obtains curves displaying
qualitatively identical behaviors. For example, for $N=50$, we find
$\mu_{crit}\simeq 0.00312\cdot M$ for $1\leq M\leq 25$.
A typical recording of the time evolution of {\it all\/} elevators
is shown in Fig.~\ref{fig2}, which displays simultaneously the movement 
of all 10 elevators as a function of time. Each line in this graph represents
the actual trajectory followed by an individual elevator. The 10
lines are not individually discernible in Fig.~\ref{fig2}: for $\mu=0.01$
because the 10 elevators move in an uncorrelated way with their
individual trajectories  significantly overlapping each other, 
and for higher $\mu$ values because they then synchronize. 

The point of the figure is to show that for
$\mu-$values of the order of $\simeq0.1$ the elevators move always 
stochastically, 
{\it i.e.}~independently of each other, reflecting essentially 
the randomness of the arrival of calling passengers in each floor.
As $\mu$ increases one sees the appearance of white intervals along 
the lowest level. 
Such  intervals  represent {\it quiescent\/}
periods, when all elevators are busy elsewhere. 
Thus, as $\mu$ increases there are sequences of roughly periodic 
time intervals 
where the majority of elevators
{\it bunch\/} at the lowest quota (ground level), indicating
a relatively high  synchronization of the arrivals on the ground floor
as the number of calls increases. 

\begin{figure}[htbp]
\centerline{~~~~~~~~\psfig{file=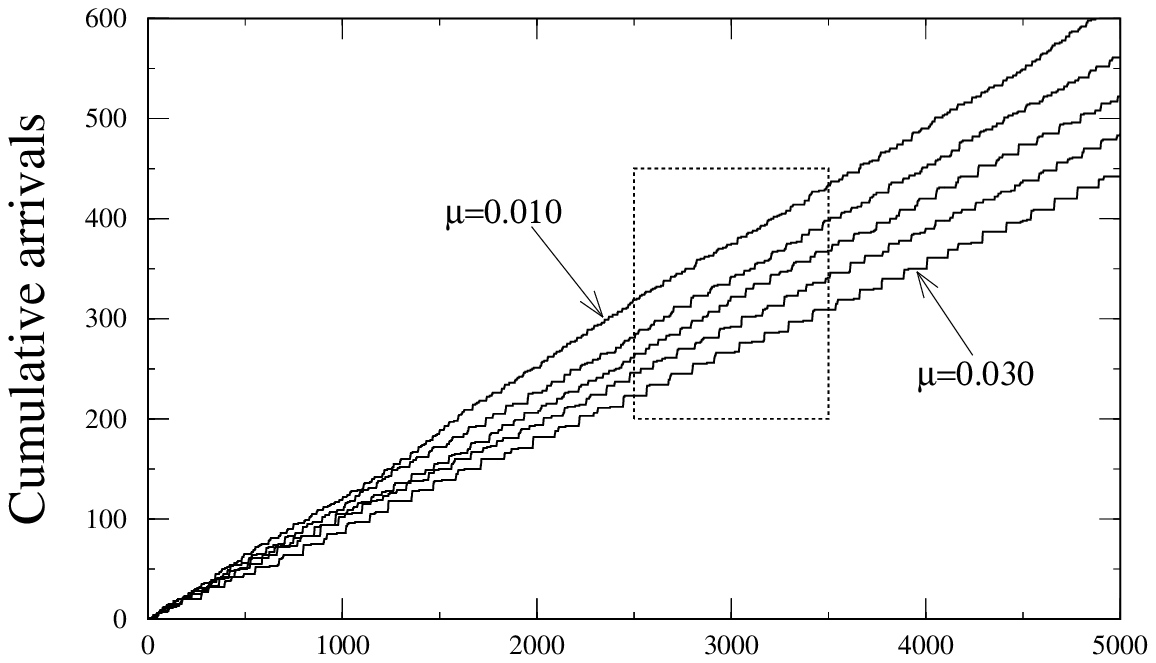,width=9cm}}
\vskip -1.7cm
\centerline{~~~~~~~~\psfig{file=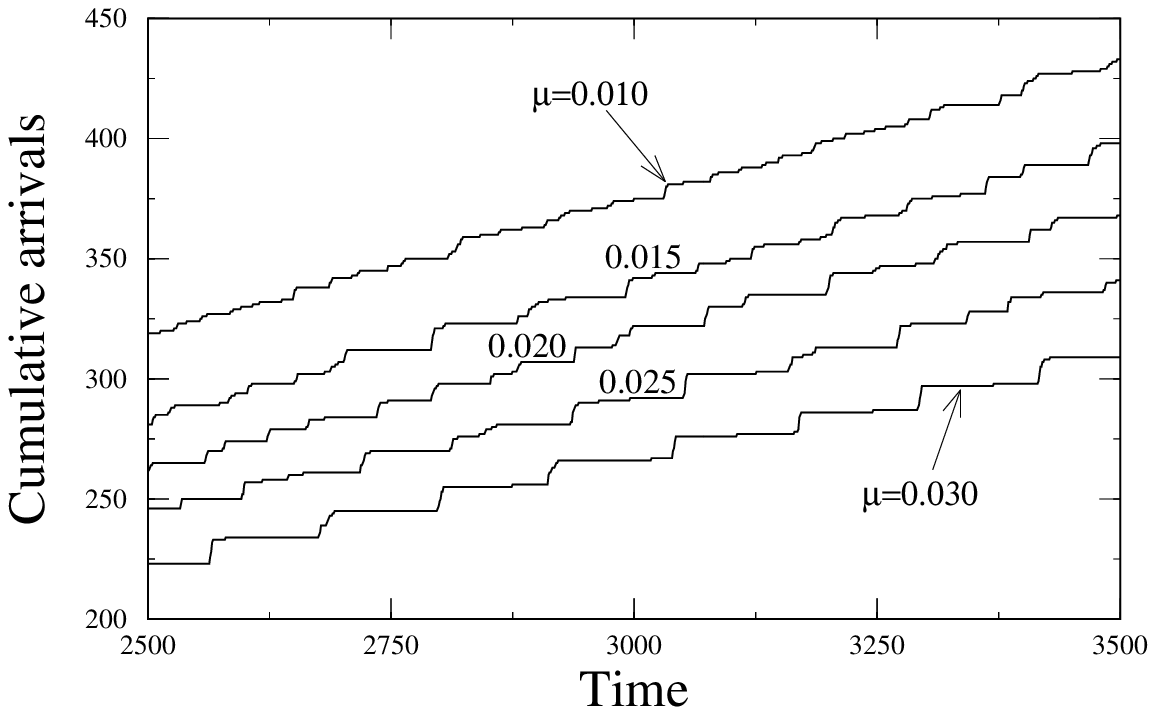,width=9cm}}
\caption{Time evolution of the cumulative number of elevators
arriving on the ground floor. The lower picture is a magnification of
the portion inside the rectangle in the upper one.}
\label{fig3}
\end{figure}

\begin{figure}[htbp]
\centerline{~~~~~~~~\psfig{file=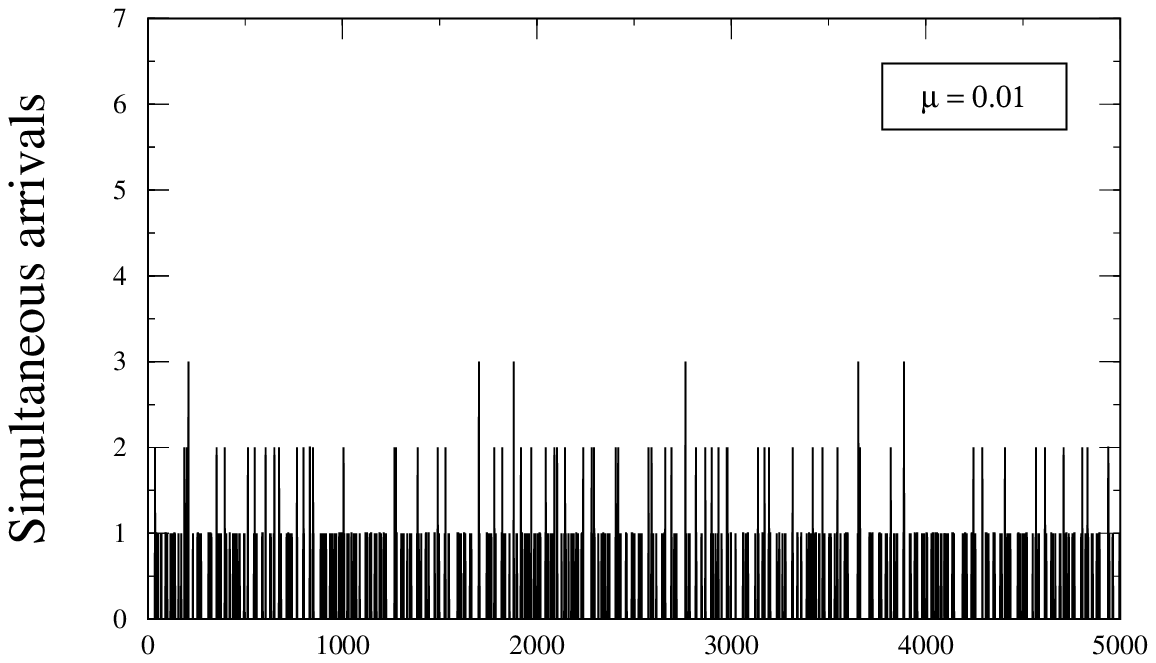,width=9cm}}
\vskip -1.65cm
\centerline{~~~~~~~~\psfig{file=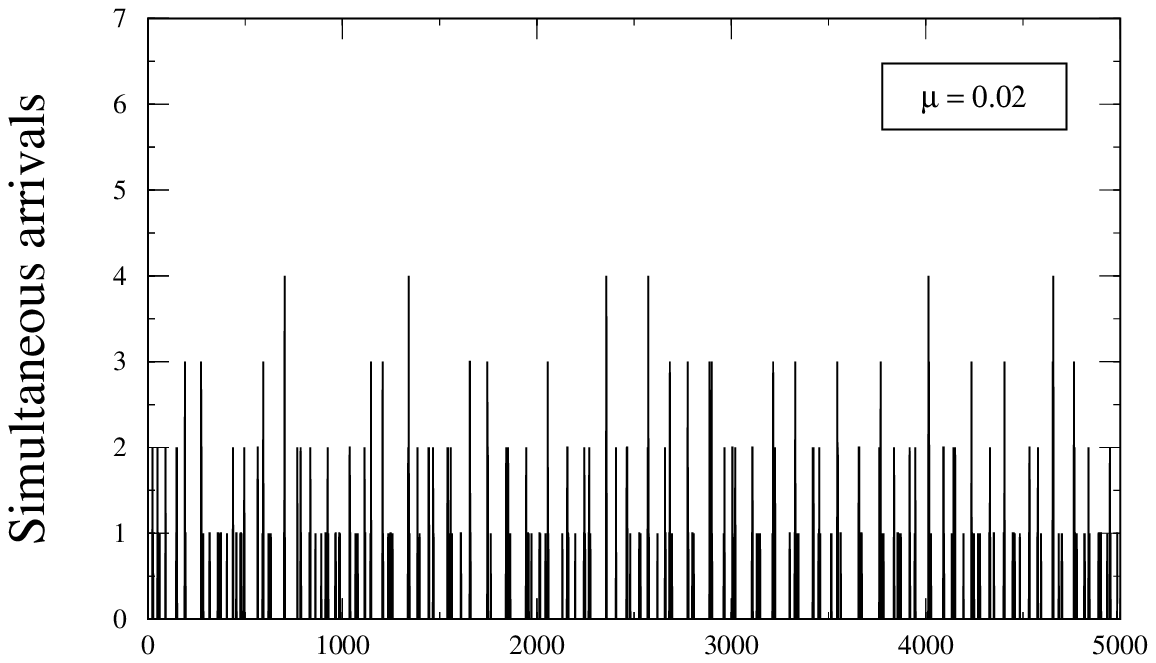,width=9cm}}
\vskip -1.65cm
\centerline{~~~~~~~~\psfig{file=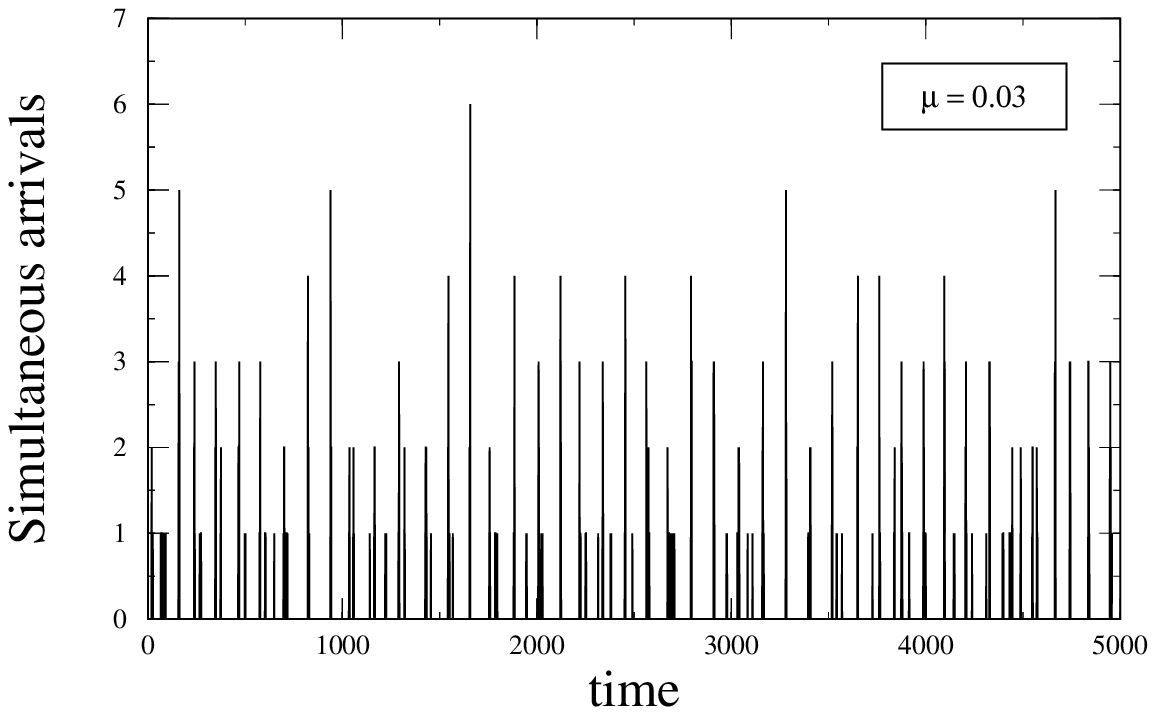,width=9cm}}
\caption{Histograms showing the number of simultaneous arrivals
as a function of time and population density.}
\label{fig4}
\end{figure}

An alternative way to recognize the synchronization is by monitoring
what happens during the time evolution of the cumulative number of 
elevators arriving on the ground floor. 
This is recorded  in Fig.~\ref{fig3}.
For small population densities one observes essentially a linear 
increase. For higher densities the number of
arrivals increases discontinuously, roughly periodically, 
via ``quantum'' jumps.
These increases occur in the
relatively short time intervals when several elevators arrive
simultaneously. The smallest (and barely visible) vertical jump 
corresponding to the arrival of a {\it single\/} elevator may be more easily
recognized from the curve corresponding to the lowest density, 
{i.e.}~$\mu=0.01$.
The synchronization and its corresponding characteristic frequency may
be also recognized from Fig.~\ref{fig4}. As a function of time, this 
figure shows the number of coincidental arrivals at the ground floor
for three different values of $\mu$. For low $\mu$, the most salient
feature is that elevators tend to arrived independently, as indicated
by the large number of single events recorded. As $\mu$ increases,
there is a redistribution, with many more elevators arriving
simultaneously. For $\mu=0.03$, near the jamming threshold, one may
already recognize visually the presence of an underlying periodicity in
the number of simultaneous arrivals.

With respect to Fig.~\ref{fig4} it is important to notice
that the figure gives the total number of elevators arriving 
at every single clock step. This is a very stringent definition of
synchronization because it only considers as ``simultaneous'' those
arrivals happening at exactly  one  {\it single\/} time step.
A more tolerant characterization would very likely involve considering
as ``simultaneous'' all arrivals occurring within a time-window larger
than a single clock unit.
The increase in the number of quasi-simultaneous arrivals seen in
Fig.~\ref{fig4} is
responsible  for the regularly spaced jumps appearing in Fig.~\ref{fig3}.

\begin{figure}[htbp]
\centerline{~~~~~~~~\psfig{file=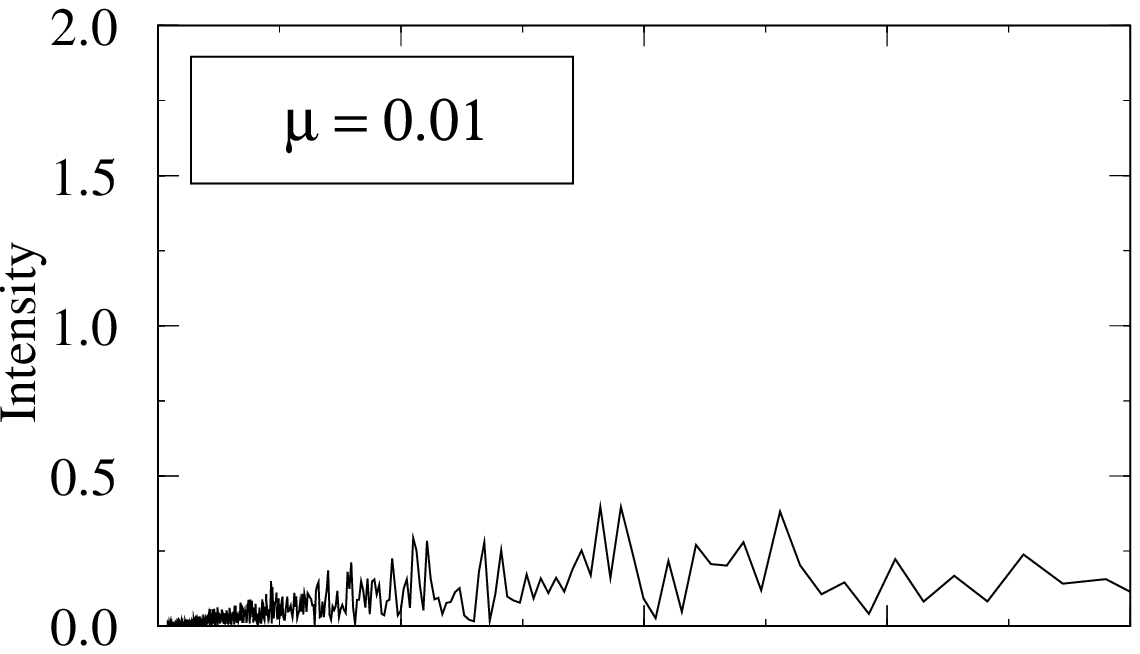,width=9cm}}
\vskip-1.63cm
\centerline{~~~~~~~~\psfig{file=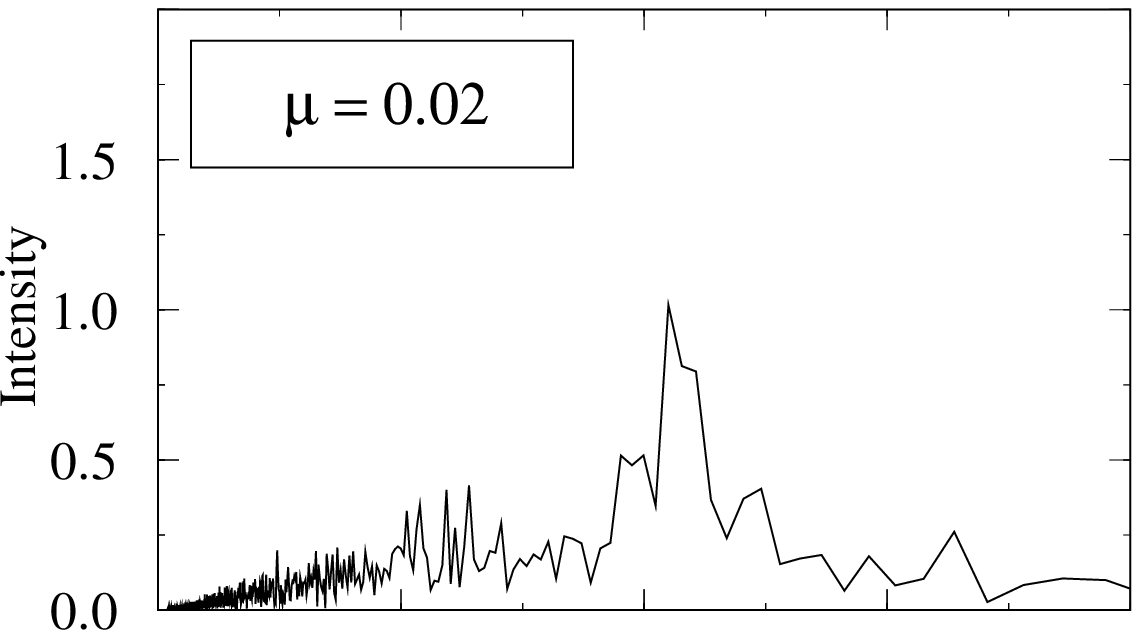,width=9cm}}
\vskip-1.63cm
\centerline{~~~~~~~~\psfig{file=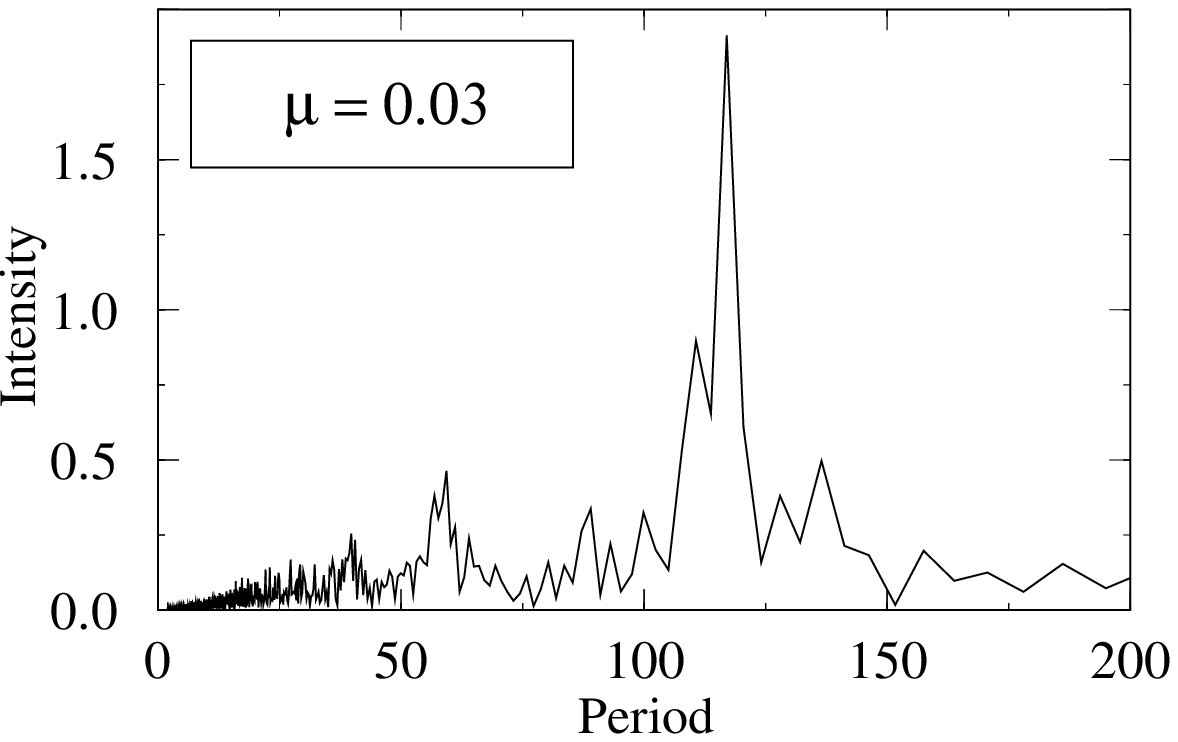,width=9cm}}
\caption{Fourier transform for  three of the  time series given in
Fig.~5, as indicated by the value of $\mu$.}
\label{fig5}
\end{figure}

To recognize quantitatively the frequencies underlying  the
regularities seen in the spacings of Fig.~\ref{fig4} we computed
Fourier transforms for the time series shown in Fig.~\ref{fig3}.
Three such transforms are given in  Fig.~\ref{fig5}, for values of
$\mu$ as indicated.

From these  figures one recognizes the appearance of a clear peak as $\mu$ 
increases, giving the  characteristic frequency with which
elevators synchronize  among themselves.

\begin{figure}[htbp]
\centerline{\psfig{file=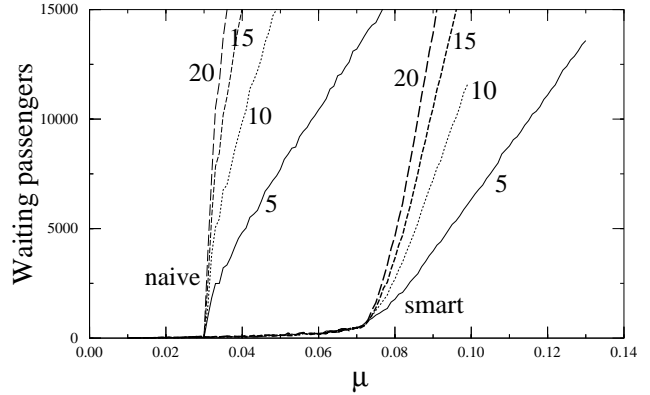,width=8.5cm}}
\caption{ Total number of waiting passengers for both {\it naive\/} 
  and {\it smart\/} elevators. Numbers refer to the elapsed  time 
after which measurements were done, in units of 1000 clock-steps.} 
\label{waiting}
\end{figure}

So far, all results presented  were obtained assuming ``naive'' elevators, 
i.e.~elevators that while going down will stop for every  call  
originating from floors below them, even after being full.
Although this is a quite common  behavior in real elevators, 
it is of interest to investigate what happens if  
elevators are smart enough to notice when they are full and use this
information to avoid unnecessary stops. 
This question is important because full elevators 
will not further contribute to the transport and their excessive stops
will only  lead to a decrease of the average speed and to a  bunching at the
highest elevator velocity (versus floor coordinate) gradient.
>From these heuristic considerations one may expect synchronization to
disappear for smart non-stopping elevators. Figure \ref{waiting} shows
the evolution of the total number of passengers waiting for elevators
as a function of $\mu$ for both types of elevators.

\begin{figure}[htbp]
\centerline{~~~~~~~\psfig{file=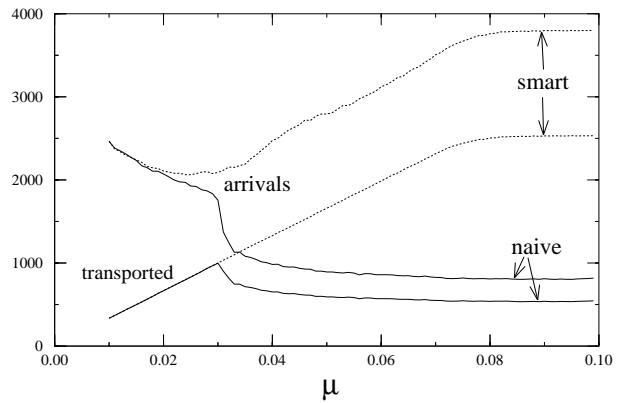,width=10cm}}
\caption{ Total number of transported passengers and total number of
  arrivals at the ground floor for both rules,  naive and smart,
  measured after 5000 clock steps. }
\label{mausi}
\end{figure}

The four sets of curves display the total of passengers waiting for
elevators after the indicated number of clock steps, divided by 1000.
It is clear that beyond $\mu_{crit}$ the transport is not efficient
anymore and the number of passengers will diverge if one waits long enough.
These curves are given to indicated the how this divergence sets in
four both types of elevators. Notice the differences in the behavior
around $\mu_{crit}$. The differences may be also recognized from
Fig.~\ref{mausi}, which shows on a single plot two sets of curves:
the total number of passengers transported after 5000 clock units and
the total number of arrivals at the bottom of the building.
As expected, one sees that smart elevators have larger $\mu_{crit}$,
being therefore  able to provide a much more efficient transport. 
We also made plots similar to Fig.~\ref{fig3} to check that
synchronization is still happening. The figure obtained is similar to 
Fig.~\ref{fig3} with smart  elevators show synchronization
over a very small  $\mu$ interval.

One interesting aspect of the simulation is the way in which the
jamming of passengers occurs. While the asymptotic number of
transported passengers is clearly limited by the number of elevators and
their capacity, it is
somewhat surprising to find that as $\mu$ increases, this limit is
attained after an  ``overshoot'' of the number of transported passengers
occurring exactly  at the jamming threshold.  
As clearly seen from Fig.~\ref{fig1},
rather than converging monotonically to the saturation level
dictated by the number and capacity of the elevators,
the most efficient transport occurs around the jamming
threshold where one finds a marked discontinuity in the derivative
of the number of transported passengers as a function of $\mu$.
Beyond the jamming threshold the  simulations 
 indicate the existence of two basic behaviors
while  passengers start to accumulate in all floors,
waiting for elevators that tend more and more to appear full. 
Immediately after the jamming threshold there is a ``relaxation interval''
of  $\mu$ values characterized by negative derivatives. After this
regime the number of transported passengers is independent of $\mu$.
All in all, the best compromise from the point of view of the
passengers wanting to leave the building is to
plan  the building to operate at the jamming threshold. 

\acknowledgments
   We thank Profs.~Werner Ebeling and Kunihiko Kaneko for helpful 
   discussions and for their kind interest in this work.

\end{multicols}
\end{document}